\documentclass[prl,twocolumn,superscriptaddress,showpacs]{revtex4}

\usepackage{graphicx,color}
\usepackage{amsmath}
\usepackage{amssymb}

\newcommand{\Tr}{\mathop{\text{Tr}}\nolimits}

\newcommand\beq            {\begin{equation}}
\newcommand\bea           {\begin{equation}\begin{array}l\displaystyle}
\newcommand\eeq            {\end{equation}}
\newcommand\bes           {\begin{subequations}}
\newcommand\esu           {\end{subequations}}

\begin{document}

\title{Simulation of two-flavors symmetry-locking phases in ultracold fermionic mixtures}

\author{Luca Lepori}
\affiliation{Departament de F\'{i}sica, Universitat Aut\`{o}noma de Barcelona, E-08193 Bellaterra, Spain.}
\affiliation{IPCMS (UMR 7504) and ISIS (UMR 7006), Universit\'{e} de Strasbourg and CNRS, Strasbourg, France.}

\author{Andrea Trombettoni}
\affiliation{CNR-IOM DEMOCRITOS Simulation Center, Via Bonomea 265, 
I-34136 Trieste, Italy}
\affiliation{SISSA and INFN, Sezione di Trieste,
via Bonomea 265, I-34136 Trieste, Italy.}

\author{Walter Vinci}
\affiliation{London Centre for Nanotechnology and Computer Science, University College London,
17-19 Gordon Street, London, WC1H 0AH, United Kingdom.}
\affiliation{University of Pisa, Department of Physics "E. Fermi" and 
INFN, Pisa, Italy.}

\begin{abstract}
We describe an ultracold fermionic set-up  
where it is possible to synthesize a superfluid phase 
with symmetry obtained by locking independent  
invariance groups of the normal state. 
In this phase, named  two-flavors symmetry-locking phase (TFSL), non-Abelian fractional vortices 
with semi-integer flux and gapless non-Abelian Goldstone 
modes localized on them appear. Considerations on the possible 
experimental realization of the TFSL are also provided.
\end{abstract}


\maketitle

\begin{figure}[h!]
\begin{center}
\includegraphics[scale=0.35]{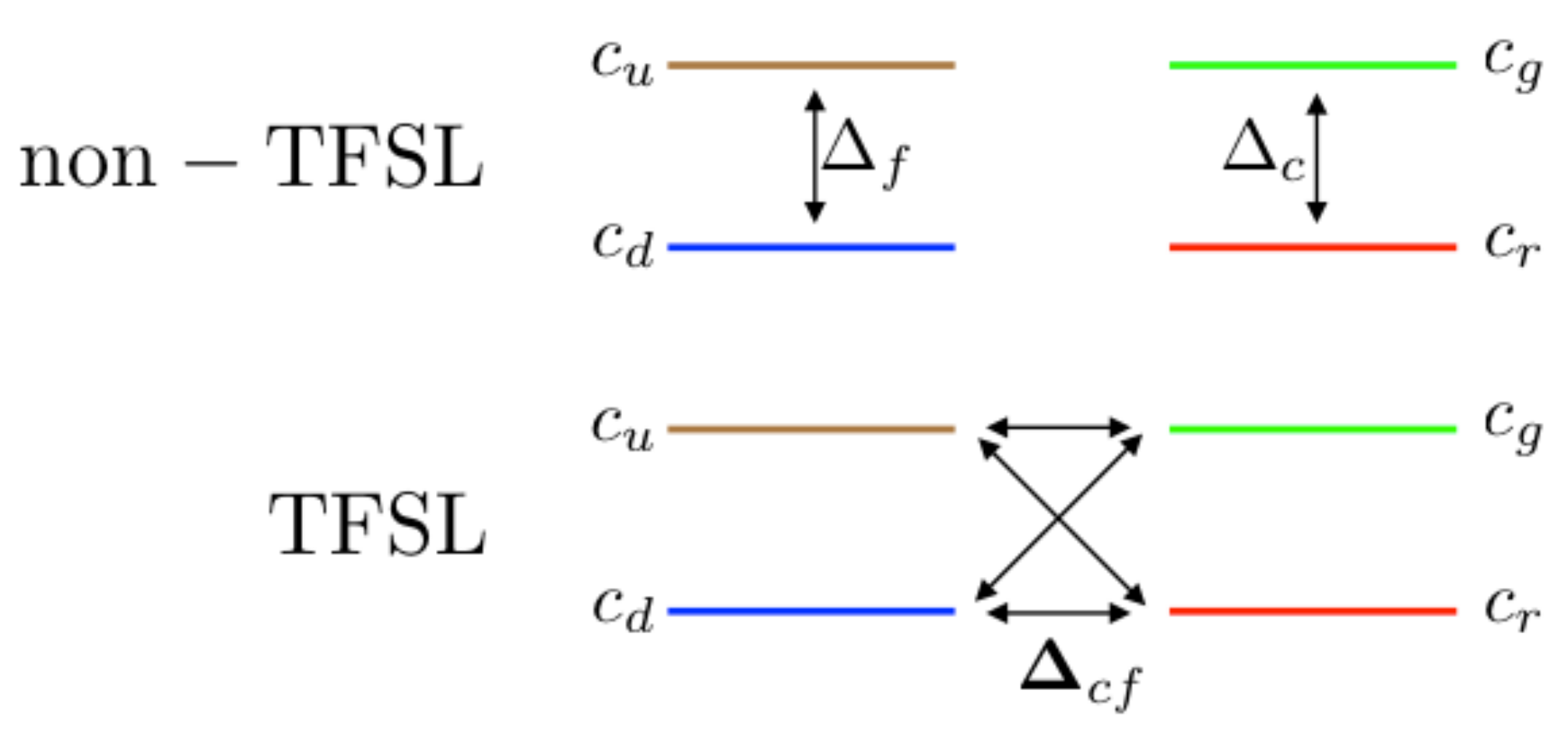}
\caption{Schematic representation of the pairings in the non-TFSL phase 
(top) and in the TFSL phase 
(bottom).}  
\label{fig1}
\end{center}
\end{figure}

{\bf Introduction --}
Trapped ultracold atoms provide an excellent tool to simulate 
strongly interacting quantum systems~\cite{bloch08}.
The main reason is the high level of tunability and the
very precise measurements achievable on such systems. 
Their versatility is considerably enlarged by two further ingredients: 
optical lattices~\cite{LibroAnna} and gauge potentials~\cite{dalibard11}. 
The possibility to synthesize Abelian and non-Abelian 
gauge fields, also in the presence of optical 
lattices~\cite{jaksch03,osterloh05,aidelsburger12,jimenez12,hauke12,atala14}, 
promises a better understanding of an increasing number of relevant
physical systems and new phases of gauge field theories.\\ 
In the non-Abelian case, hyperfine levels of suitable atoms are 
typically used as 
internal degrees of freedom where the gauge potential is defined: an 
advantageous proposal is given by earth-alkaline atoms 
\cite{gerbier10}.  
To date, only static gauge fields have been experimentally simulated, but 
proposals for dynamical fields recently appeared in literature 
\cite{Dyn1,Dyn2,Dyn3,Dyn4,Dyn5}.\\
A further promising and challenging application of ultracold atoms 
and synthetic gauge fields is the emulation of relativistic models 
relevant to high energy physics: recent proposals focused 
on 2D~\cite{Dirac2D,guineas,wu08,Juzeliunas,lim08,hou09,lee09,alba13} and 
3D~\cite{Lamata,toolbox,LMT,RevCold} Dirac fermions, 
Wilson fermions and axions~\cite{toolbox}, 
neutrino oscillations~\cite{neutosc} and 
extra dimensions~\cite{extra}. Important experimental achievements in 
this direction are  the recent synthesis of 2D Dirac fermions ~\cite{Tarr}
and the Haldane model \cite{Haldane} on honeycomb-like lattices.\\
These intensive efforts are also expected to progress 
towards a better understanding of strongly coupled non-Abelian gauge theories, 
like quantum chromodynamics (QCD), and potentially to give new insights 
on the QCD phase diagram. Indeed in QCD 
various problems stay unsolved, like the origin of color 
confinement~\cite{dualat}, chiral symmetry breaking (CSB),
as well as the dynamics of nucleons, nuclear matter or quarks under extreme
conditions~\cite{Raj,Manna}. In the light of the developments 
mentioned above, ultracold atoms can become a precious tool in 
the investigation of these important topics.\\
A very important concept arising in high-energy physics is 
the locking of symmetries, meaning that independent symmetries (global or local) of the Hamiltonian
 are mixed when the system enters in a certain phase, generally superfluid.
A prominent example of this mechanism is 
the color-flavor locking (CFL),
characterized by a mixing of  
the (local) $SU(3)_c$ color and the (global) 
$SU(3)_f$ flavor symmetries~\cite{Alf}, independent in the normal state.
 A CFL regime is predicted 
at very large densities, as in the core of ultra-dense neutron stars, 
where the huge pressure and temperature allow quarks 
to be deconfined~\cite{Raj,Manna}. 
Similar superfluid phases are also relevant in the strongly 
coupled regime, at the CSB transition and especially
in the context of dual superconductivity models 
of color confinement~\cite{Dual1,Dual2}. Dual models are supported by 
lattice simulations~\cite{dualat} and by exactly 
solvable supersymmetric theories~\cite{SeiWit}, where exact dualities and 
correspondences can be established between confining phases and weakly 
coupled CFL phases~\cite{Carlino:2000uk}.\\
The characterizing property of a symmetry-locked phase is the presence of a 
complex pattern of spontaneous breaking of non-Abelian symmetries 
(SSB) induced by a multicomponent superfluid condensate. 
This is at the heart of its remarkable properties:   
the appearance of vortices and monopoles 
with semi-integer flux and non-Abelian degrees of freedom confined on 
them~\cite{Auzzi:2003fs,Shifman:2004dr,Hanany:2003hp,Eto:2005yh,Auzzi:2004if},
and the breaking of translational invariance with the presence of
ordered structures (as crystals and nets)~\cite{Raj,Manna}. 
Vortices in CFL phases also play an important role in determining the 
magnetic behaviour of neutron stars~\cite{Balach,NittanonAb,Vinci:2012mc} and 
the properties of confinement and CSB \cite{Eto:2009wu}.\\
We describe here a realistic proposal for the synthesis 
of a superfluid phase with locking between two global symmetries. 
Our relatively simple set-up differs 
from a realistic quantum simulation of QCD in two aspects: 
{\em i)} QCD fermions have both color and flavor quantum numbers 
at the same time; {\em ii)} QCD has both local (color) 
and global (flavor) symmetries, while in our model both the
symmetries are global. Indeed we name the obtained symmetry-locked phase
 as two flavors symmetries locking phase (TFSL).
However, concerning {\em i)} 
our simplified model is still able to capture 
the effects of the various symmetries and 
their spontaneous breaking, independently from the internal space 
they are realized on. Similarly, concerning {\em ii)}, in real QCD 
some important phenomena 
are related to the breaking of global symmetries, as the CSB transition 
(described effectively by the ungauged Nambu-Jona-Lasinio model  \cite{NJL}) 
close the deconfinement point and the consequent appearance 
of scalar mesons \cite{dualat, Wei2}.
Moreover, the experimental implementation of the scheme proposed here 
provides a first and (in our opinion) necessary 
step towards the quantum simulation of a CFL phase 
featuring symmetry breaking of local symmetries, for which the use of 
dynamical gauge fields for ultracold atoms is required.\\
Another motivation for the implementation of a symmetry-locked phase 
with ultracold fermionic mixtures as proposed here is that 
-- even with only global symmetries -- 
non-Abelian fractional vortices (NAFV) appear. Remarkably the NAFV studied in this paper differ 
from the previously observed ones \cite{revUeda,Nitta1,stamp} 
in the origin of their non-Abelianity and fractionally. 
As a consequence their braiding properties are also different. 
The simulation of TFSL and CFL phases by using ultracold atoms would 
be then important to detect and study such exotic objects, 
especially due to the possibility 
of directly measuring  various correlation functions 
(including ones involving the density operator)~\cite{fetter09}.\\
{\bf The model --}
We consider an optical lattice loaded with a mixture 
of four ultra-cold fermionic species: e.g., two different atoms, 
each one trapped in two hyperfine levels. 
We conventionally denote the four types of 
fermions by two different quantum labels (each one spanning two values) 
$c=\{r,g\}$ and $f=\{u,d\}$.
The $c$ and $f$ index are in turn collectively denoted as \emph{flavors}
since, in the model that we are going to adopt, each of them is associated with a \emph{global}
non-Abelian symmetry.\\
We assume for simplicity that the considered Fermi mixture 
is loaded in an optical lattice 
(say cubic), but the main conclusions of the paper 
and the occurrence of the TFSL do not depend on the presence of the lattice.\\
We assume an Hubbard Hamiltonian of the form 
$\hat{H}=\hat{H}_{kin}+\hat{H}_{int}$~\cite{hofstetter02},
where $\hat{H}_{kin}$ is the usual tight-binding hopping Hamiltonian 
(the hopping coefficient $t$ between nearest-neighbor sites is assumed for 
simplicity to be independent of $c$ and $f$ indices)
and 
\begin{equation}
\begin{array}{c}
\hat{H}_{int} = - U_c  \sum_{i,c \neq c^{\prime}} {n}_{i;c}  {n}_{i;c^{\prime}}  
- U_f  \sum_{i,f \neq f^{\prime}} {n}_{i;f} {n}_{i;f^{\prime}} -\\
{}\\
- U_{cf} \sum_{i,c,f} {n}_{i;c} {n}_{i;f},
\label{Hint}
\end{array}
\end{equation}
where $n_{i,\sigma}={c}_{i;\sigma}^{\dag} {c}_{i;\sigma}$ 
is the number operator of the fermion $\sigma=r,g,u,d$ on the site $i$. 
We also consider balanced mixtures: the atomic numbers for the $c$ and $f$ species  
coincide ($N_c=N_f$ with $N_g=N_r=N_c/2$ and 
$N_u=N_d=N_f/2$). In (\ref{Hint}) 
it is $U_{c} = - \frac{4 \pi  \hbar^2}{m} a_{c} \int d\vec{r} \, \phi_{c}^4$ and   
$U_{f} = - \frac{4 \pi  \hbar^2}{m} a_{f} \int d\vec{r} \, \phi_{f}^4$, 
where $a_c$ and $a_f$ are respectively the $s$-wave scattering lengths 
between two $c$  and $f$ fermions
(the $\phi$ 's are the appropriate Wannier functions). Similarly 
$-U_{cf}$ is proportional to the 
$s$-wave scattering length $a_{cf}$ 
between one $c $ fermion and a $f$ one. 
The Hamiltonian in Eq.~(\ref{Hint}) has an $U(2)_{c} \times U(2)_{f} \equiv G$ symmetry enlarged to $U(4)$ if $U_{c}=U_{f}=U_{cf}$.\\
{\bf Non-TFSL and TFSL phases --} 
We discuss for the moment the simplest model, where attraction 
arises between all the fermions: $U_c,U_f,U_{cf}>0$. Moreover 
we set for simplicity $U_c = U_f \equiv U$.  Repulsive 
intra-$c$ or intra-$f$ interactions ($U_c<0$ or $U_f<0$) are 
detrimental for the occurrence of the TFSL phase: 
for instance this is  the case for ${}^{171}$Yb -$^{173}$Yb mixtures. We will discuss 
in detail this situation later, proposing solutions to obtain still the TFSL phase.\\
For $U>0$  and $U_{cf}>0$ the possible SF phases 
are parametrized by the pairings
\begin{equation}
\langle c_{i;\,u} c_{i;\,d}\rangle \equiv \Delta_f, \, \,   
\langle c_{i;\,r} c_{i;\,g}\rangle \equiv \Delta_c, \, \, 
\langle c_{i;\,c} c_{i;\,f}\rangle \equiv {\bf \Delta}_{cf} \, ,
\label{ord_par}
\end{equation}
with ${\Delta}_{c}$, ${\Delta}_{f}$ and ${\bf \Delta}_{cf}$ 
being respectively 
two generic complex numbers and a $2 \times 2$ complex matrix. 
We derive the free energy $F$ in  the
mean-field approximation, similarly to the case of 
two-component mixtures~\cite{Annett}. It is convenient 
to define the order parameter $\Delta_0$ as 
$2|\Delta_0|^2=|\Delta_c|^2+|\Delta_f|^2$ 
and to introduce the $U(2)_{c+f}$ invariants 
$\Delta_{+}^{2}=\Tr\left( {\bf \Delta}_{cf}^{\dagger} {\bf \Delta}_{cf}
\right)$,   
$\Delta_{-}^{2}=2 \det {\bf \Delta}_{cf}$.\\
The minimization of $F$ with respect to $\Delta_{\pm}$ and $\Delta_0$ gives
$|\Delta_{+}|=|\Delta_{-}|$ and $|\Delta_{c}|=|\Delta_{f}|$. We find that 
for $U_{cf} \neq U$ the gap equations are not consistent 
if both $\Delta_{+}$ and $\Delta_{0}$ are non-zero 
both $T=0$ and finite temperature and two phases are found as follows 
[see Fig.~(\ref{fig1})]:
\begin{itemize}
\item[{\em i)}] Non-TFSL phase: for $U_{cf} < U$ it is $\Delta_{+} = 0$ and 
$\Delta_{0}\neq 0$; the gap equation at $T=0$ reads: 
$\sum_k \frac{1}{\sqrt{\epsilon_k^2+|\Delta_0|^2/2}}=\frac{1}{2U}$ 
(being $\epsilon_k=\varepsilon_k-\mu$, with $\varepsilon_k$ 
the single-particle energy 
spectrum of $H_{kin}$ and $\mu$ the chemical potential); 
\item[{\em ii)}] TFSL phase: for $U_{cf} > U$ it is $\Delta_{0} = 0$ and 
$\Delta_{+}\neq 0$; the gap equation at $T=0$ reads: 
$\sum_k \frac{1}{\sqrt{\epsilon_k^2+|\Delta_+|^2/2}}=\frac{1}{2U_{cf}}$. 
\end{itemize}
Notice that $|\Delta_{+}| \neq 0$ corresponds 
to a TFSL phase, irrespectively from the value of $\Delta_c$ and 
$\Delta_f$~\cite{note_equal} .\\
The findings in \emph{i)} and \emph{ii)} are consistent with the results 
discussed in literature (e.g., see~\cite{Annett,minpar}) 
that show how many-species superfluidity tends to avoid configurations 
with multiple pairings having different symmetries and competing with each others.
The gap equations and the equation for the particle number (not written here) 
are also the same in both the TFSL phase and the non-TFSL phase, but with 
$\Delta_{+}$ and $U_{cf}$ instead of $\Delta_0$ and $U$ (both at vanishing 
and at finite temperature). These equations, as well as the corresponding equations for the chemical potentials
$\mu_c$ and $\mu_f$,
have the same functional form of the Leggett's equations for a two-component Fermi mixture across the 
BCS-BEC crossover~\cite{Zwerger}. Indeed they coincide for the diagonal pairing   ${\bf \Delta}_{cf} = \Delta \, \delta_{cf}$.
\\
{\bf Properties of the TFSL phase --} 
The key difference between the TFSL and the non-TFSL phase 
lies in the residual non-Abelian symmetry of the TFSL phase.
In fact, the non-TFSL pairings induce a SSB of the Abelian factors: 
$U(2)_c \times U(2)_f \to SU(2)_c \times SU(2)_f$. At variance, 
the TFSL pairing ${\bf \Delta}_{cf}$ satisfying $|\Delta_{-}|=|\Delta_{+}| \neq 0$ 
induces the following SSB pattern $G \to H$:
\begin{equation}
U(2)_c \times U(2)_f =  U(2)_{c+f} \times U(2)_{c-f} \to U(2)_{c + f}.
\label{pattern1}
\end{equation}
This means that the SF phase 
(as well as ${\bf \Delta}_{cf}$) has a residual symmetry group 
$H = U(2)_{c + f}$ given by the set of elements 
$({\cal U}_c, {\cal U}_f) = ({\cal U}_{c}, {\cal U}_{c}^{-1}) = 
({\cal U}_{f}^{-1}, {\cal U}_{f})$, ${\cal U}_c$ and ${\cal U}_f$ 
belonging respectively to $U(2)_{c}$ and $U(2)_{f}$. 
We define also 
$U(2)_{c - f} = ({\cal U}_{c}, {\cal U}_{c}) = ({\cal U}_{f}, {\cal U}_{f})$.
$U(2)_{c + f}$, acts on ${\bf \Delta}_{cf}$ as 
${\cal U}_{c+f}  \, {\bf \Delta}_{cf} \, {\cal U}_{c+f}^{-1}$, 
and thus involves at the same time $c$ and $f$ transformations, 
originally independent. 
This mechanism is called symmetry locking~\cite{Alf}.\\
The most general form of ${\bf \Delta}_{cf}$ 
compatible with the gap equations for the TFSL phase is 
${\bf \Delta}_{cf} = \Delta \, \tilde{{\cal U}}_c^{\dagger} \, 
\tilde{{\cal U}}_f$,  with 
$\tilde{{\cal U}}_c$ and $\tilde{{\cal U}}_f$ 
also belonging to $U(2)_{c}$ and $U(2)_{f}$. This form parametrizes the coset $G/H$.
Without any loss of generality, we can always 
perform a symmetry transformation to put the gap matrix into a diagonal form:  
${\bf \Delta}_{cf} = \Delta \, \delta_{cf} \equiv \Delta \, {\bf I}$, 
with $\delta_{cf}$ being the Kronecker delta,  or ${\bf \Delta}_{cf} = \Delta \, \sigma_x$ ($\sigma_i$ are as usual the Pauli matrices) \cite{notasigma}.\\
The SSB at the heart of TFSL 
implies the existence (both with and without vortices) of 
non-Abelian gapless Goldstone modes propagating 
in the whole condensate and described by the coset $G/H \sim S^2$. 
Such modes are related to spatial and time fluctuations 
of the order parameter $\Delta_{cf}$ in the general set described above. 
This is analogous to the presence 
of spin waves in the B phase of superfluid $^3$He~\cite{Volovik:2003fe}.
The spectrum of gapless excitations 
is  completely different from the non-TFSL scenario, 
where only two Abelian Goldstone modes appear.\\
{\bf Comments about experimental feasibility--}  
An instance of experimental setup realizing our scheme is provided (possibly without optical lattice, in the continuous) by $^{171}$Yb and $^{173}$Yb mixtures~\cite{Dic,Yip}, the first one having a $1/2$ 
hyperfine degeneracy and the second one a $5/2$ multiplet - 
in the latter case  only two levels have to be  populated selectively.
The main motivation for this choice is that these atoms have 
natural interspecies interactions not depending on the particular 
hyperfine levels considered \cite{alkaline}. 
Even if this property may hold in other fermionic mixtures (for instance involving earth-alkaline atoms), at the best of our
knowledge $^{171}$Yb -$^{173}$Yb is at the present time the unique stable mixture having it experimentally realized.\\ 
The $c$ index indicates now the two hyperfine levels of $^{171}$Yb, while 
$f$ refers to two hyperfine levels of $^{173}$Yb.\\
The scattering lengths are respectively
$a_{171} \equiv a_c =-3 \, a_{0}$, $a_{173} \equiv a_{f}=+200 \, a_{0}$ and 
$a_{171-173} \equiv a_{cf}=-578 \, a_{0}$~\cite{Tc} ($a_0$ being the Bohr radius). 
Therefore $U_c$ is positive 
and very small, $U_{cf}$ is positive and  $U_f$ is negative and relatively 
large (corresponding to the sensible repulsion between $f$ atoms).\\
The main effect of the intraspecies
interactions is to induce an effective 
imbalance between the chemical potentials of the two atomic isotopes: 
this competes with the formation of the TFSL state 
driven by the dominating interaction $\propto a_{171-173}$. 
In the mean-field Hamiltonian from (\ref{Hint}) 
(and the same populations for all the hyperfine 
levels both of $^{171}\text{Yb}$ 
and $^{173}\text{Yb}$), this imbalance results in a term  
$U_c \, \nu_c \sum_{i,c} {c}_{i;\,c}^{\dag} {c}_{i;\,c} + U_f \, 
\nu_f \sum_{i,f} {c}_{i;\,f}^{\dag} {c}_{i;\,f}$, being $\nu_c = \nu_f \equiv \nu$ 
the filling of each atomic species. 
In this way, a different phase instead of TFSL can be realized. While in this paper we are not strictly interested to analyze the phase diagram 
of $\text{Yb}$ mixtures, we address briefly how to overcome this problem.
Tuning the interactions by optical Feshbach resonance is presently 
not a realistic solution, due to significant atomic losses.\\
The first possibility is to unbalance the $c$ and $f$ populations in number, $N_c > N_f$,
such that the difference of the Fermi energies $\Delta E _F \equiv E^{(c)}_F-E^{(f)}_F = 2 t \, (\mathrm{cos} \, k^{(c)}_F - \mathrm{cos} \, k^{(f)}_F)$ in the normal states  
approaches the quantity  $(U_f \, \nu_f - U_c \, \nu_c)$. In this way the two unbalances compensates each others,
allowing the appearance of the TFSL phase.
An alternative  (but experimentally more difficult) way to create a compensating unbalance is to allow a difference 
in the hopping terms, $(t_c - t_f) > 0$.\\
A different possibility involves two additional species in the $f$ multiplet and it assumes possible
to couple $f$-atoms in pairs (say (1 , 3) and (2 , 4)) via two Raman pulses with amplitude $\Omega$.
This model has a $U(2)_c \times U(4)_f$ global symmetry 
that is explicitly broken to $U(2)_c \times U(2)_f$ by the Raman pulses. 
The new eigenstates
in the normal state, where the residual symmetry is exactly realized, 
are obtained introducing the operators 
$\eta_{13/24}^{(\pm)} (\vec{k}) = \frac{c_{1/2} \pm c_{3/4}}{\sqrt{2}} (\vec{k})$ 
with energy $\varepsilon_{\pm}(\vec{k}) = \varepsilon(\vec{k}) 
\pm \Omega$. 
A direct mean-field calculation
shows that tuning $\Omega \approx (U_f - U_c) \, \nu$ 
compensates the original effective imbalance between 
the $^{171}\text{Yb}$ levels and $\eta_{13/24}^{(-)}$ \cite{Pethick} 
and allows for the TFSL pairing between them. 
Instead the states $\eta_{13/24}^{(-)}$, largely 
(anyway more than the unmixed levels of $^{173}\text{Yb}$) 
imbalanced with respect of $^{171}\text{Yb}$, stay unpaired. 
This system then formally reduces to the simple model 
discussed in Eq.~(\ref{Hint}), and consequently the occurrence and 
detection of TFSL and non-TFSL phases proceeds in the same way.
We observe that this second approach, relying on the use of a selective Raman coupling among pairs of atoms, may be challenging for $^{171}$Yb -$^{173}$Yb mixtures. However it could be applied to other fermionic mixtures where the population unbalance might be difficult to be realized.\\
{\bf Experimental detection --} 
From the previous discussion of their pairing structures, 
it emerges how to experimentally 
discriminate between TFSL and non-TFSL superfluids. In the latter situation
one finds two conventional superfluids 
with non-vanishing pairings $\Delta_c$ and $\Delta_f$, while 
in the TFSL phase a pairing between $c$ and $f$ fermions arises.
in principle along all the possibilities allowed by the TFSL symmetry: 
${\bf \Delta}_{cf} = \Delta \, \tilde{{\cal U}}_c^{\dagger} \, 
\tilde{{\cal U}}_f$
[see a schematic representation in Fig.~\ref{fig1}]. 
Therefore by observing the interference patterns 
 created by the fermionic condensates 
below the critical temperature it is possible to 
discriminate the non-TFSL and TFSL phases (since 
the order parameters are respectively proportionals to ${\bf I}$
and to $\sigma_x$).\\
Similarly, inducing vortices (e.g., by rotation~
\cite{fetter09}) 
and measuring the density profiles one can immediately distinguish the two 
phases. \\
 {\bf Non-Abelian fractional vortices --}
In this Section we study the behavior of the TFSL phase under the application of a synthetic Abelian
magnetic field $\vec{B}_{synth}$ along $\hat{z}$: 
this can be obtained by putting under rotation 
the lattice~\cite{Cooper} or by optical means~\cite{lin09}, 
for example with the implementation of multipod schemes~\cite{dalibard11}. 
When $\vec{B}_{synth}$ is applied to the non-TFSL phase, vortices emerge in the spatial configuration 
of $r$-$g$ and $u$-$d$ condensates, while in the TFSL phase there is a locked 
spatial configuration involving both $c$ and $f$ atoms. The detection 
of vortices and the related spatial distributions of the fermionic species provides a 
first clear-cut characterization of the TFSL phase.\\
Vortices in the TFSL phase exhibit two remarkable 
properties: {\em a)} they host localized non-Abelian Goldstone modes (NAGM); 
{\em b)} they have fractional flux. To see the first property we 
start from an SF phase described at 
$\vec{B}_{synth} = 0$ by  ${\bf \Delta}_{cf} = \Delta \, {\bf I}$. 
At $\vec{B}_{synth} \neq 0$, the energetically favored configurations 
host vortices in the condensates $\Delta_{r u}$ and $\Delta_{g d}$.
In the following we assume that the vortices in the two components do not overlap completely.  
This implies that ${\bf \Delta }_{cf} ({\vec r})$ is not  any longer  a multiple of the identity matrix. Indeed let us consider  the spatial dependence of the gap parameter 
around a vortex configuration, for example in the  $\Delta_{r u}$ component:
\begin{equation}
{\bf \Delta }_{cf} (\vec{r})= \Delta({r}) \, ({\bf I} + 
\sigma_z \, \eta_z ({r})  ) \,e^{\frac{i}{2} \, ({\bf I} + \sigma_z) \, \theta} \, ,
\label{pbreak}
\end{equation}
where $\theta =  [0, 2 \pi )$ is the spatial angle around the vortex and $r$ is the distance.  
Moreover we have  $\eta_z(0)=1$ at the core of the vortex while  $\Delta({r}) \to \Delta$ and 
$\eta_z({r}) \to 0$ at distances ${r}$ large compared to the vortex typical size. \\
Eq.~(\ref{pbreak}) implies that the TFSL phase 
undergoes an additional SSB $H  \to H_V$
along the following pattern:
\begin{equation}
U(2)_{c + f} \to \big(U(1)_{c+f} \times U(1)_{\sigma^z} \big)/Z_2 \, ,
\label{pattern}
\end{equation}
where $U(1)_{\sigma^z}$ (generated by $\sigma^z$) is  contained in  $SU(2)_{c + f}$.
Two additional Goldstone modes then appear around the positions $\vec{r} \approx \vec{r_i}$ of the separated vortices,
where $\eta_z\neq 0$.
These modes can be made manifest by 
noticing that, once a solution of the type shown in Eq.~(\ref{pbreak}) is found, 
one can generate  a continuous family of  degenerate solutions 
by applying $c$-$f$ rotations:
\begin{equation}
{\cal U}_{c+f}  \, {\bf \Delta}_{cf}  ({r})\, {\cal U}_{c+f}^{-1}= 
\Delta({r}) \, ({\bf I} + 
\vec {\cal S} \cdot \vec \sigma \, \eta_z ({r})  ) \,e^{\frac{i}{2} \, ({\bf I} + 
\vec {\cal S} \cdot \vec \sigma  ) \, \theta} \, ,
\label{family}
\end{equation}
with $\vec \sigma = (\sigma_x,\sigma_y,\sigma_z)$ and 
$|\vec {\cal S}|=1$ 
being a normalized vector parameterizing the  Goldstone modes. 
The discussion above implies that vortices in the TFSL phase are  endowed with NAGM forming a non linear representation of the group $H$~\cite{Wei2} and spanning the target space 
$H/H_V=S^2$.\\
A comparison with vortices in the non-TFSL phase helps clarification. 
In that case the lowest-energy configuration  is
\begin{equation}
\Delta_{\alpha} (\vec{r})= \tilde{\Delta}_{\alpha}(r) \, e^{i  \, \theta} \, ,
\label{pbreak0}
\end{equation}
with $\alpha = c,f$: these vortices are Abelian~\cite{Balach,Manton},
since (\ref{pbreak0}) parameterizes $\frac{U(2)_{c,f}}{SU(2)_{c,f}} \equiv U(1)_{c,f}$. Conversely, a totally abelian configuration as in (\ref{pbreak0}) has finite energy also in the TFSL phase but this energy is bigger than for (\ref{family}), then the latter one is selected.\\
We comment now our assumption of vortex separation in the TFSL case. 
The issue can be tackled
by explicit calculation of the Landau-Ginzburg functional (LGF) (not reported here) and via direct comparison to the LGF results for multi-component mixtures~\cite{revUeda}. 
It turns out that vortices in the different components of ${\bf \Delta}_{cf}$ generally repel each others~\cite{NittaInt1,NittaInt2},\cite{noterepel}.\\ 
{\bf Fractional flux --} Remarkably, fundamental  vortices in the TFSL phase 
have a fractional (half)  flux compared to an abelian vortex of a non-TFSL phase, 
as can be seen by a direct computation or 
by more general arguments~\cite{Auzzi:2003fs}.
Indeed, vortices  are classified via their quanta of flux
by the element of the first homotopy group $\pi_1$~\cite{Coleman, Nak} defined on the SSB pattern (\ref{pattern1}), that is 
$\pi_1 \left(\frac{U(2)_{c+f} \times U(2)_{c-f}}{U(2)_{c+f}} \right) \sim \pi_1 \big(U(2)_{c-f} \big) = \mathbb{Z}/\mathbb{Z}_2$ in a TFSL phase and 
$\pi_1(U(1)_{c,f}) = \mathbb{Z}$ in a non-TFSL one. \\
The discussed properties lead us to refer to the obtained solitons as 
NAFV~\cite{note3}.
Various types of fractional vortices have already been studied 
in various inhomogeneous systems, e.g. 
in Josephson junctions systems~\cite{Jos1,Jos2,Jos3}, 
in spin-1 Bose condensates~\cite{revUeda} and Helium $3$~\cite{Bosesp1}. 
However fractionality in these cases generally 
requires  $\Delta (\vec{r})$ to be asymptotically $\propto e^{i \kappa \theta}$, 
with $\kappa=p/q$ a rational number: 
this implies in turn nontrivial braiding properties due 
to the presence of spatial branch-cuts in the definition of $\Delta (\vec{r})$ 
\cite{Ivanov}. Conversely the braiding is trivial in our case since it involves no anyonic statistics  (notice the integer numbers in front of $\theta$ in the phases appearing in (\ref{pbreak}) and (\ref{family})). 
Other examples of fractional vortices but with integer 
phase 
arise in multi-components Bose superfluids~\cite{twocomp} 
or metallic liquid hydrogen~\cite{Bab}.
There an explicit interaction between the pairings is needed, 
unlike the present case, where
this property is possible only thanks 
to the group structure of $U(2)_{c-f}$.\\
{\bf Outlook --} 
We proposed and discussed a set-up of a fermionic ultracold mixture 
in which it is possible to synthesize two-flavors locked (TFSL) phases. 
Due to their symmetry, these TFSL phases host exotic non-Abelian 
vortices with semi-integer flux and localized gapless 
modes confined on them. The origin of the non-Abelianity, the braiding properties,  
the mechanism and the consequences of fractionality for such 
vortices are discussed.
To the best of our knowledge, 
we predicted for the first time the existence of such solitons 
in an experimentally accessible set-up.\\ The effect 
of repulsive intra-species interactions has been discussed, 
showing that they can destroy the TFSL phase:  two different solutions have been proposed,  
based on the creation of a counter-unbalance compensating this effect.
 A discussion of the detection of the TFSL and non-TFSL phases
has been also provided.\\
A partial list of important developments to the present work concern: a)
the derivation of the Ginzburg-Landau 
coefficients including corrections beyond  mean-field 
(as done for the two-species attractive Hubbard model in~\cite{Iazzi});
b) the study of interactions between the vortices and their spatial configurations, especially far from the critical point,   
by  solving self consistently the Bogoliubov-De Gennes equations~\cite{BdG} on certain vortex backgrounds; c)
the application of non-Abelian gauge potentials,
confirming a radically different response to these potentials by non-TFSL and TFSL phases;
d)
a more accurate simulation of the CFL phase of  high-density QCD,  by the inclusion of dynamical 
(local) symmetries; e)  the realization of non translationally invariant SF colored phases 
\cite{Raj,Manna} and f) the synthesis of chiral symmetry breaking phases 
via superfluidity in semimetallic systems~\cite{Lamata,toolbox,LMT}. 
Furthermore, it would be highly interesting to extend our results in presence 
of $p$-wave pairings. This  would give raise to non-Abelian Majorana fermions 
of the type discussed in~\cite{yasui11}, relevant for topological quantum computation~\cite{Nayak}.\\

{\bf Acknowledgements:} 
The authors are pleased to thank E. Babaev, M. Baranov, L. Fallani, 
A. Gorshkov, M. Mannarelli, M. Nitta and G. Sierra for fruitful discussions.
L.L. acknowledges a grant from Banco de Santander and 
financial support from ERDF.


\begin{thebibliography}{XX}

\bibitem{bloch08}
I. Bloch, J. Dalibard, and W. Zwerger, 
Rev. Mod. Phys. {\bf 80}, 885 (2008).

\bibitem{LibroAnna}
M. Lewenstein, A. Sanpera, and V. Ahufinger, 
{\em Ultracold atoms in optical lattices: simulating quantum many-body systems} 
(Oxford, Oxford University Press, 2012).

\bibitem{dalibard11}
J. Dalibard, F. Gerbier, G. Juzeli{\=u}nas, and P. \"Ohberg, 
Rev. Mod. Phys. {\bf 83}, 1523 (2011). 

\bibitem{jaksch03}
D. Jaksch and P. Zoller, New Journ. Phys. {\bf 5}, 56 (2003). 

\bibitem{osterloh05}
K. Osterloh, M. Baig, L. Santos, P. Zoller, and M. Lewenstein, 
Phys. Rev. Lett. {\bf 95}, 010403 (2005).

\bibitem{aidelsburger12}
M. Aidelsburger, M. Atala, S. Nascimb\`ene, S. Trotzky, 
Y.-A. Chen, and I. Bloch, Phys. Rev. Lett. {\bf 107}, 255301 (2011). 

\bibitem{jimenez12}
K. Jimenez-Garcia, L. J. Le Blanc, R. A. Williams, 
M. C. Beeler, A. R. Perry, and I. B. Spielman,  
Phys. Rev. Lett. {\bf 108}, 225303 (2012).

\bibitem{hauke12}
P. Hauke, O. Tieleman, A. Celi, C. \"Olschl\"ager, J. Simonet, 
J. Struck, M. Weinberg, P. Windpassinger, K. Sengstock, 
M. Lewenstein, and A. Eckardt, Phys. Rev. Lett. {\bf 109}, 145301 (2012).

\bibitem{atala14}
M. Atala, M. Aidelsburger, M. Lohse, J. T. Barreiro, B. Paredes, 
and I. Bloch, \verb|arXiv:1402.0819|

\bibitem{gerbier10}
F. Gerbier and J. Dalibard, New Journ. Phys. {\bf 12}, 033007 (2010).

\bibitem{Dyn1}
E. Zohar, J. I. Cirac, and B. Reznik,  
Phys. Rev. Lett. {\bf 109}, 125302 (2012); 
{\em ibid.} {\bf 110}, 055302 (2013); {\em ibid.} {\bf 110}, 125304 (2013).

\bibitem{Dyn2}
D. Banerjee, M. Dalmonte, M. M\"uller, E. Rico, P. Stebler, 
U.-J. Wiese, and P. Zoller, Phys. Rev. Lett. {\bf 109}, 175302 (2012).

\bibitem{Dyn3}
L. Tagliacozzo, A. Celi, A. Zamora, and M. Lewenstein, Ann. Phys. 
{\bf 330}, 160 (2013).

\bibitem{Dyn4}
D. Banerjee, M. B\"ogli, M. Dalmonte, E. Rico, P. Stebler, U.-J. Wiese, 
and P. Zoller, Phys. Rev. Lett. {\bf 110}, 125303 (2013). 

\bibitem{Dyn5}
M. J. Edmonds, M. Valiente, G. Juzeli{\=u}nas, L. Santos, and P. \"Ohberg, 
Phys. Rev. Lett. {\bf 110}, 085301 (2013).

\bibitem{Dirac2D}
S.-L. Zhu, B. Wang, and L.-M. Duan, Phys. Rev. Lett. {\bf 98}, 260402 (2007).

\bibitem{guineas}
B. Wunsch, F. Guinea, and F. Sols, 
New Journ. Phys. {\bf 10}, 103027 (2008).

\bibitem{wu08}
C. Wu and S. Das Sarma, 
Phys. Rev. B {\bf 77}, 235107 (2008). 

\bibitem{Juzeliunas}
G. Juzeli{\=u}nas, J. Ruseckas, M. Lindberg, L. Santos, and P. \"Ohberg, 
Phys. Rev. A {\bf 77}, 011802(R) (2008).

\bibitem{lim08}
L.-K. Lim, C. M. Smith, and A. Hemmerich, 
Phys. Rev. Lett. {\bf 100}, 130402 (2008).

\bibitem{hou09}
J.-M. Hou, W.-X. Yang, and X.-J. Liu, Phys. Rev. A {\bf 79}, 043621 (2009).

\bibitem{lee09}
K. L. Lee, B. Gr\'emaud, R. Han, B.-G. Englert, and C. Miniatura, 
Phys. Rev. A {\bf 80}, 043411 (2009).

\bibitem{alba13} 
E. Alba, X. Fernandez-Gonzalvo, J. Mur-Petit, J. J. Garcia-Ripoll, 
and J. K. Pachos, Ann. Phys. {\bf 328}, 64 (2013).

\bibitem{Lamata}
L. Lamata, J. L\'eon, T. Sch\"atz, and E. Solano, 
Phys. Rev. Lett. {\bf 98},  253005 (2007).

\bibitem{toolbox}
A. Bermudez, L. Mazza, M. Rizzi, N. Goldman, M. Lewenstein, and M. A. 
Martin-Delgado, Phys. Rev. Lett. {\bf 105}, 190404 (2010).

\bibitem{LMT}
L. Lepori, G. Mussardo, and A. Trombettoni, Europhys. Lett. {\bf 92}, 
50003 (2010).

\bibitem{RevCold}
L. Mazza, A. Bermudez,  N. Goldman, M. Rizzi, M. A. Martin-Delgado, and 
M. Lewenstein, New. Journ. Phys. {\bf 14}, 01500 (2012).

\bibitem{neutosc}
Z. Lan, A. Celi, W. Lu, P. \"Ohberg, and M. Lewenstein, 
Phys. Rev. Lett. {\bf 107}, 253001 (2011).

\bibitem{extra}
O. Boada, A. Celi, J. I. Latorre, and M. Lewenstein, 
Phys. Rev. Lett. {\bf 108}, 133001 (2012). 

\bibitem{Tarr} 
L. Tarruell, D. Greif, T. Uehlinger, 
G. Jotzu, and T. Esslinger, Nature {\bf 483}, 302 (2012).

 

\bibitem{Haldane}
G. Jotzu, M. Messer,
R. Desbusquois,
M. Lebrat,
T. Uehlinger, D. Greif, and T. Esslinger,
arXiv: 1406.7874.

\bibitem{dualat}
H. J. Rothe, {\em Lattice gauge fields: an introduction} 
(Singapore, World Scientific, 2005).

\bibitem{Raj}
M. Alford, A. Schmitt, K. Rajagopal, and T. Sch\"afer, 
Rev. Mod. Phys. {\bf 80}, 1455 (2008).

\bibitem{Manna}
R. Anglani, R. Casalbuoni, M. Ciminale, R. Gatto, N. Ippolito, 
M. Mannarelli, and M. Ruggieri, Rev. Mod. Phys. {\bf86}, 509  (2014). 


\bibitem{Alf}
M. Alford, K. Rajagopal, and F. Wilczek, Nucl. Phys. B {\bf 537}, 443 (1999).

\bibitem{Dual1}
S. Mandelstam, Phys. Rep. {\bf 23 C}, 245 (1976).

\bibitem{Dual2}
G. 't Hooft, Nucl. Phys. B {\bf 190}, 455 (1981).

\bibitem{SeiWit}
N. Seiberg and E. Witten, Nucl. Phys. B {\bf 426}, 19 (1994).

\bibitem{Carlino:2000uk} 
G. Carlino, K. Konishi, and H. Murayama, 
Nucl. Phys. B {\bf 590}, 37 (2000).
  
\bibitem{Auzzi:2003fs} 
R. Auzzi, S. Bolognesi, J. Evslin, K. Konishi, and A. Yung, 
Nucl. Phys. B {\bf 673}, 187 (2003).

\bibitem{Auzzi:2004if} 
R. Auzzi, S. Bolognesi, J. Evslin, K. Konishi, and H. Murayama, 
Nucl. Phys. B {\bf 701}, 207 (2004).

\bibitem{Shifman:2004dr} 
M. Shifman and A. Yung, 
Phys. Rev. D {\bf 70}, 045004 (2004).
  
\bibitem{Hanany:2003hp} 
A. Hanany and D. Tong, Journ. High En. Phys. {\bf 07}, 037 (2003).
  
\bibitem{Eto:2005yh} 
M. Eto, Y. Isozumi, M. Nitta, K. Ohashi, and N. Sakai, 
Phys. Rev. Lett. {\bf 96}, 161601 (2006).

\bibitem{Balach}
A. P. Balachandran, S. Digal, and T. Matsuura, 
Phys. Rev. D {\bf 73}, 074009 (2006).

\bibitem{NittanonAb}
E. Nakano, M. Nitta, and T. Matsuura
Phys. Rev. D {\bf 78}, 045002 (2008).

\bibitem{Vinci:2012mc} 
W. Vinci, M. Cipriani and M. Nitta,
Phys. Rev. D {\bf 86}, 085018 (2012).

\bibitem{Eto:2009wu} 
M. Eto, E. Nakano, and M. Nitta, 
Nucl. Phys. B {\bf 821}, 129 (2009).

\bibitem{NJL}
Y. Nambu and G. Jona-Lasinio, 
Phys. Rev. {\bf 122}, 345 (1961). 

\bibitem{Wei2}
S. Weinberg, {\em The quantum theory of fields}, Vol. 2 (Cambridge, 
Cambridge University Press, 1996).

\bibitem{revUeda} 
C. Kasamatsu, M. Tsubota, and M. Ueda, Int. Journ. Mod. Phys. B 
{\bf 19}, 1835 (2005).

\bibitem{Nitta1}
M. Kobayashi, Y. Kawaguchi, M. Nitta, and M. Ueda, 
Phys. Rev. Lett. {\bf 103}, 115301 (2009).

\bibitem{stamp}
D. M. Stamper-Kurn and M. Ueda, 
Rev. Mod. Phys. {\bf 85}, 1191 (2013).

\bibitem{fetter09}
A. L. Fetter,
Rev. Mod. Phys. {\bf 81}, 647 (2009). 

\bibitem{hofstetter02}
W. Hofstetter, J. I. Cirac, P. Zoller, E. Demler, and M. D. Lukin, 
Phys. Rev. Lett. {\bf 89}, 220407 (2002). 




\bibitem{Annett}
J. F. Annett, {\em Superconductivity, superfluids, and condensates} 
(Oxford, Oxford University Press, 2004), Chaps. 6-7.

\bibitem{note_equal}
For $U=U_{cf}$ beyond mean-field effects should be considered 
to study the existence and the properties of the TFSL phase.

\bibitem{minpar}
C. Honerkamp and W. Hofstetter, 
Phys. Rev. Lett. {\bf 92}, 170403 (2004);
Phys. Rev. B. {\bf 70}, 094521 (2004).

\bibitem{Zwerger}
{\em The BCS-BEC crossover and unitary Fermi gas}, W. Zwerger ed. 
(Heidelberg, Springer, 2012). 

\bibitem{notasigma}
To see the $U(2)_{c+f}$ invariance when $\Delta_{cf} = \Delta \, \sigma_x$, it suffices to consider the following choice for the flavor transformations $U_f = \sigma_x^\dagger \, U_c  \, \sigma_x$, with $U_c$ chosen freely.  It immediately follows that $\sigma_x = U_c \, \sigma_x U_f^{-1}$.

\bibitem{Volovik:2003fe} 
G. E. Volovik, {\em The universe in a helium droplet} 
(Oxford, Oxford University Press, 2003).



\bibitem{Dic}
D. B. M. Dickerscheid, Y. Kawaguchi, and M. Ueda, 
Phys. Rev. A {\bf 77}, 053605 (2008).

\bibitem{Yip}
S.-K. Yip, Phys. Rev A {\bf 83}, 063607 (2011).




\bibitem{alkaline}
A. V. Gorshkov, M. Hermele, V. Gurarie, C. Xu, P. S. Julienne, J. Ye, P. Zoller, E. Demler, M. D. Lukin and A. M. Rey,
Nature Phys. 6, 289 - 295 (2010).




\bibitem{Tc}
S. Taie, Y. Takasu, S. Sugawa, R. Yamazaki, T. Tsujimoto, R. 
Murakami, and Y. Takahashi, 
Phys. Rev. Lett. {\bf 105}, 190401 (2010).






\bibitem{Pethick}
C. J. Pethick and H. Smith, {\em Bose-Einstein Condensation in Diluite Gases}, 
2nd ed., Chap. 16, Cambridge University Press (2008).

\bibitem{Cooper}
N. R. Cooper, Adv. Phys. {\bf 57}, 539 (2008).

\bibitem{lin09}
Y. J. Lin, R. L. Compton, K. Jimenez-Garcia, J. V. Porto, and 
I. B. Spielman, Nature {\bf 462}, 628 (2009).

\bibitem{Manton}
N. Manton and P. Sutcliffe, {\em Topological solitons} (Cambridge, 
Cambridge University Press, 2004).




\bibitem{NittaInt1}
E. Nakano, M. Nitta, and T. Matsuura, Phys. Lett. B {\bf 672}-1, 61 (2009). 

\bibitem{NittaInt2}
M. Eto, K. Kasamatsu, M. Nitta, H. Takeuchi, and M. Tsubota,
Phys. Rev. A {\bf 83}-6, 063603 (2011).


\bibitem{noterepel}
The only exceptions are the diagonal cases ${\bf \Delta}_{cf} = \Delta  \, {\bf I}$ and ${\bf \Delta}_{cf} = \sigma_x$, where no mean field interaction terms 
arise between vortices in different condensates.  In this scenario
a more precise evaluation of LGF coefficients, taking into account quantum fluctuations, is required. However, even if not spontaneous, vortex separation can be driven artificially here in various ways, for instance by an additional soft perturbation $\vec{B}_{synth}^z \propto \hat z \,\sigma_z$  in the flavour space. This field induces an unbalance in the number of vortices in $\Delta_{r u}$ and $\Delta_{g d}$, hence vortex separation.


\bibitem{Coleman}  S. Coleman, {\em The magnetic monopole fifty years later}, in 
{\em The unity of the fundamental interactions}, A. Zichichi ed. 
(London, Plenum, 1983).

\bibitem{Nak}
M. Nakahara, {\em Geometry, topology and physics}, 2nd ed. (Bristol, 
Institute of Physics, 2003).




\bibitem{note3}
High-energy physics community refers to solitons like ours simply as
non-Abelian vortices~\cite{Auzzi:2003fs,Balach}, since the patterns labeled 
by $\pi_1(U(2)_{c-f})$ are partly in the non-Abelian group $SU(2)_{c-f}$. 
We added here the
word "fractional" to avoid confusion with another 
more common type of non-Abelian vortex
in condensed matter literature. This definition is in fact mostly used 
to denote the non-Abelianity of the first homotopy group $\pi_1$ 
defining the set of vortex charges: this feature critically affects 
the behavior of the vortices under merging and braiding 
\cite{Nitta1,stamp}. 

\bibitem{Jos1}
A. Ustinov, Appl. Phys. Lett. {\bf 80}, 3153 (2002). 

\bibitem{Jos2}
E. Goldobin, A. Sterck, T. Gaber, D. Koelle, and R. Kleiner, 
Phys. Rev. Lett. {\bf 92}, 057005 (2004).

\bibitem{Jos3}
J. Pfeiffer, M. Schuster, A. A. Abdumalikov, Jr., and A. V. Ustinov, 
Phys. Rev. Lett. {\bf 96}, 034103 (2006). 

\bibitem{Bosesp1}
D. Vollhardt and P. Woelfle, {\em The superfluid phases of Helium 3} 
(London, Taylor \& Francis, 1990).

\bibitem{Ivanov}
D. A. Ivanov, Phys. Rev. Lett. {\bf 86}, 268 (2001).

\bibitem{twocomp}
E. Babaev, Phys. Rev. Lett. {\bf 89}, 067001 (2002).

\bibitem{Bab}
E. Babaev and N. W. Ashcroft, Nature Phys. {\bf 3}, 530 (2007).



\bibitem{Iazzi}
M. Iazzi, S. Fantoni, and A. Trombettoni, Europhys. Lett. {\bf 100}, 36007 
(2012).

\bibitem{BdG}
P.-G. de Gennes,  \emph{Superconductivity Of Metals And Alloys}, Advanced Book Program (Perseus Books, 1999).


\bibitem{yasui11}
S. Yasui, K. Itakura, and M. Nitta,
Phys. Rev. D {\bf 81}, 105003 (2010);
Phys. Rev. B {\bf 83}, 134518 (2011).

\bibitem{Nayak}
C. Nayak, S. H. Simon, A. Stern, M. Freedman, S. Das Sarma
Rev. Mod. Phys. {\bf 80}, 1083 (2008).


\end{thebibliography}
\end{document}